\documentclass[english,prb,aps,amsmath,amssymb,reprint,superscriptaddress,showpacs]{revtex4-2}
\usepackage[T1]{fontenc}
\usepackage[utf8]{inputenc}
\usepackage{color}
\usepackage{babel}
\usepackage{array}
\usepackage{multirow}
\usepackage{amstext}
\usepackage{graphicx}
\usepackage[unicode=true,
 bookmarks=true,bookmarksnumbered=false,bookmarksopen=false,
 breaklinks=true,pdfborder={0 0 0},pdfborderstyle={},backref=false,colorlinks=true,pdfpagemode=FullScreen]
 {hyperref}
\hypersetup{pdftitle={Si(111) under tensile strain: evidence for the c(2x4) surface reconstruction},
 pdfauthor={R.A. Zhachuk, A.V. Latyshev, J. Coutinho},
 pdfsubject={68.35.bg, 68.35.Gy, 68.35.Md},
 pdfkeywords={Germanium, Diffuion, Surface reconstruction, Strain, Ab initio calculations},
 allcolors=blue}

\makeatletter

\providecommand{\tabularnewline}{\\}

\makeatother

\begin{document}
\title{Impact of strain and surface reconstruction on long-range diffusion
of Ge atoms on Ge$(111)$ surface}
\author{R. A. Zhachuk}
\email{zhachuk@gmail.com}

\affiliation{Institute of Semiconductor Physics, pr. Lavrentyeva 13, Novosibirsk
630090, Russia}
\author{A. V. Latyshev}
\affiliation{Institute of Semiconductor Physics, pr. Lavrentyeva 13, Novosibirsk
630090, Russia}
\author{J. Coutinho}
\affiliation{I3N, Departament of Physics, University of Aveiro, Campus Santiago,
3810-193 Aveiro, Portugal}
\date{\today}
\begin{abstract}
We investigate the effect of surface reconstruction and strain on
diffusion of adsorbed Ge atoms on Ge$(111)\textrm{-}5\times5$ and
Ge$(111)\textrm{-}7\times7$ surfaces by means of first principles
calculations. Stable adsorption sites, their energies, diffusion paths,
and corresponding activation barriers are reported. We demonstrate
that the decisive migration path is located near the corner holes
of surface structures, and they are associated with formation of weak
bonds between the adsorbed Ge atom and surface dimers (within the
$5\times5$ or $7\times7$ structures). The results show that Ge diffusion
rates on $5\times5$ and $7\times7$ reconstructed Ge$(111)$ surfaces
should be similar. Conversely, the diffusion barrier on a compressively
strained Ge$(111)$ surface is considerably higher than that on a
strain-free surface, thus explaining previous experimental results.
Comparable diffusion rates on $5\times5$ and $7\times7$ reconstructed
surfaces are explained by the identical local atomic arrangements
of these structures. The increase of the migration barrier on a strained
surface is explained by dimer bond strengthening upon surface compression,
along with a weakening of bonds between the adsorbed Ge and dimer
atoms.

\noindent \href{https://doi.org/10.1103/PhysRevB.107.245305}{DOI:10.1103/PhysRevB.107.245305}
\end{abstract}
\maketitle

\section{INTRODUCTION}

Surface diffusion is an important topic with direct impact on thin
film growth, its morphology and nanostructure formation. Besides temperature,
several parameters can be used to change the surface diffusivity.
For example, it can be influenced by the surface atomic structure,
substrate chemical composition, type of diffusing species or surface
strain.

The strain dependence of atomic diffusion on metal surfaces, where
bonds are omnidirectional, is relatively well understood. Surface
compressive strain pushes the diffusing atoms away from the crystal
surface so that they experience a less corrugated surface potential.
As a result, the energy barrier for the metallic systems is found
to decrease with increasing compressive strain \citep{bru95,rat97}.

On the other hand, strain and structure dependence of diffusion on
semiconductor surfaces can be more complex because of the directional
and localized nature of bonds between surface and adsorbed atoms.
For instance, strong diffusion anisotropy has been observed on Si$(001)\textrm{-}2\times1$
surface: the diffusivity parallel to dimer rows is faster than that
perpendicular to the rows by three orders of magnitude \citep{mo91,mo92,swa96}.
With increasing tensile strain, the diffusivity across the dimer rows
is further frustrated (according to both experiments and first principles
calculations \citep{shu01,zoe00}) while that along the dimer rows
is enhanced (according to calculations of Shu \emph{et~al.}~\citep{shu01})
or does not change (according to experiments of Zoethout and co-workers
\citep{zoe00}).

Experimental studies of surface diffusion on Si$(111)\textrm{-}7\times7$
also revealed an interesting phenomenon — the mobility of adsorbed
atoms inside the Si$(111)\textrm{-}7\times7$ half unit cells (HUCs)
was in many cases found to be high at room temperature, while under
the same conditions, hops between neighboring HUCs were rare \citep{zha10,uch04,sat00,gom96,vit99,uch02,mys01,pol02,uch03,cus01}.
This is somewhat unexpected since Si$(111)\textrm{-}7\times7$ HUCs
contain a high density of dangling bonds, to which adsorbed atoms
can connect strongly. The above findings were partially explained
by a first principles study showing that the high radical density
on Si$(111)\textrm{-}7\times7$ HUCs is precisely at the origin the
low migration barriers. Accordingly, concurrent breaking/formation
of closely spaced bonds within the HUCs lead to a relatively flat
potential energy surface for adatom motion \citep{cha03,cho98}.

Cherepanov and Voigtländer studied the surface diffusion of adatoms
(Ge or Si) on Ge$(111)$ and Si$(111)$ surfaces in the range $T=400\textrm{-}700\,\mathrm{K}$
by following the density of two-dimensional islands formed upon submonolayer
deposition by scanning tunneling microscopy (STM) \citep{che04,che02}.
By preparing substrates that differed by a single parameter only,
they were able to experimentally separate the influence of various
factors impacting the density of formed islands. The change of the
surface diffusion barrier was estimated from the change of island
density with help of Venables theory of nucleation \citep{Venables1994},
\emph{i.e.}, assuming typical values for the critical nucleation size
and pre-exponential factors. It was found that the most influential
factors were the substrate and deposited materials (Si or Ge). A relatively
slower diffusivity of Si (in comparison to that of Ge) was found on
both Si(111) and Ge(111) substrates, and that was attributed the formation
of stronger Si-$X$ bonds (as compared to Ge-$X$ ones), with $X=\{\textrm{Si},\textrm{Ge}\}$.
It was also shown that the change of Ge$(111)$ surface structure
from $7\times7$ to $5\times5$ had a negligible effect on the island
formation rate (surface diffusivity), and more surprisingly, the diffusion
barrier for Ge (Si) adatoms increased with increasing surface compressive
strain. The latter observations are difficult to understand and were
left unexplained \citep{che04,che02}.

The present work aims at addressing the above questions by means of
atomistic first principles calculations. We find that the migration
path on both $5\times5$ and $7\times7$ reconstructed surfaces is
located near the corner holes of the surface structure, and they involve
the formation of weak bonds between adsorbed Ge atoms and surface
dimers. It is shown that equal diffusion rates on $5\times5$ and
$7\times7$ reconstructed surfaces are due to the identical local
atomic arrangements of these structures. On the other hand, the diffusion
barrier increase under surface compression is due to strengthening
of surface dimer bonds, accompanied by weakening of bonds between
adsorbed Ge and dimer atoms at the transition state.

\section{CALCULATION DETAILS}

\begin{figure*}
\includegraphics[clip,width=14cm]{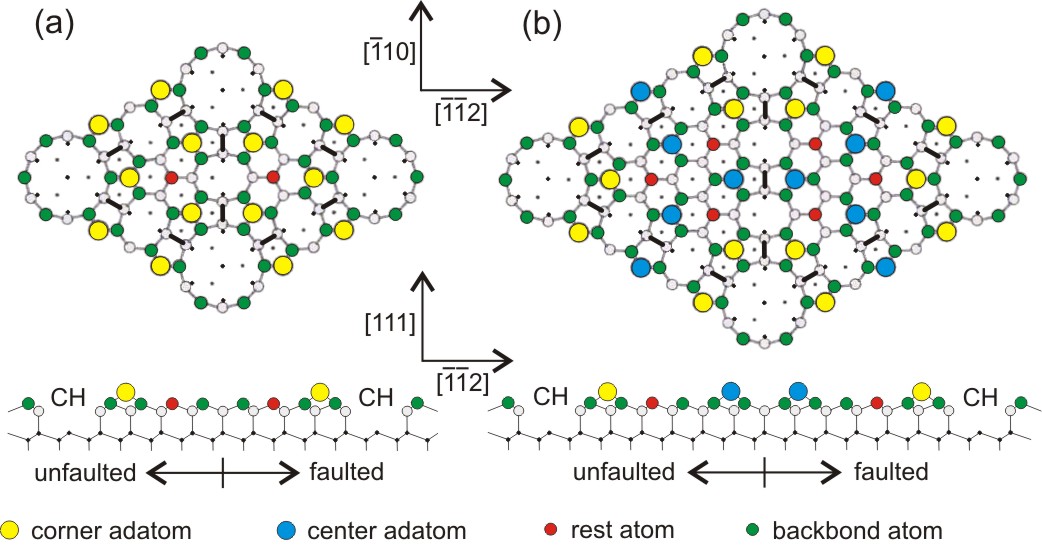}

\caption{\label{fig1} Top (upper panels) and sectional (lower panels) views
of (a) $5\times5$ and (b) $7\times7$ structures of Ge$(111)$ surface
according to the DAS model \citep{tak85}. Sectional views are in
the $(1\overline{1}0)$ plane cutting the long diagonal of the $5\times5$
and $7\times7$ surface unit cells, showing the nearest neighbor bonding
structure. Thick rods (in the top views) edging the perimeter of the
cells represent Ge-Ge dimers. CH stands for corner hole.}
\end{figure*}
The calculations were carried out using the pseudopotential \citep{tro91}
density functional theory (DFT) SIESTA code \citep{sol02,gar20} within
the generalized gradient approximation (GGA) to the exchange and correlation
interactions between electrons as parameterized by Perdew, Burke and
Ernzerhof (PBE) \citep{per96}. The Kohn–Sham wavefuctions were described
with help of linear combinations of atom centered orbitals of the
Sankey–Niklewski type, which included multiple zeta orbitals and polarization
functions \citep{sol02}.

The supercell geometries consisted of repeating slabs of four $(111)$-oriented
Ge bilayers separated by a $16$~Å wide vacuum gap along the Cartesian
\emph{z}-direction. All Ge dangling bonds at the unreconstructed bottom
surface were saturated by hydrogen atoms, and those Ge-H units were
kept frozen during atomic relaxations. Top layers of the slabs were
modified according to the dimer-adatom-stacking fault (DAS) atomic
models of $5\times5$ and $7\times7$ reconstructions \citep{tak85},
as depicted in Fig.~\ref{fig1}. According to that model, each $5\times5$
($7\times7$) unit cell contains (i) a stacking fault in one of its
HUC between 1st and 2nd bilayer, (ii) a corner hole (CH) corresponding
to one missing atom in second atomic layer (first bilayer) and, hence,
leaving a dangling bond at its center atom in the third atomic layer
(second bilayer), (iii) 6 (9) dimers forming domain walls along the
boundary of one of its HUCs, (iv) 6 (12) adatoms in $\mathrm{T_{4}}$
sites in a $2\times2$-like environment, and (v) 2 (6) rest atoms
in the first atomic layer, the dangling bonds of which are not saturated
by bonding to the adatoms. The two non-equivalent HUCs with and without
stacking faults are called faulted and unfaulted respectively.

Atoms from the two upper Ge layers of the slab and the adsorbed Ge
atom contributed with two sets of \emph{s} and \emph{p} orbitals plus
one \emph{d} orbital to describe the Kohn-Sham states. On the other
hand, Ge atoms from the two bottom-most layers of the slab had only
one set of \emph{s} and \emph{p} orbitals, and all H atoms were assigned
a single \emph{s} orbital. The electron density and potential terms
were calculated on a real space grid with spacing equivalent to a
plane-wave cut-off of $200$~Ry. For the Brillouin-zone (BZ) integration
of the $5\times5$ surface structure we used a $4\times4\times1$
$\mathbf{k}$-point grid \citep{mon76}, while for the $7\times7$
structure we used $\Gamma$-point only.

The energetics of a Ge adatom on the Ge$(111)$ surface was studied
by means of mapping the potential energy surfaces (PES) as a function
of the adatom position. PES maps were produced for positions of the
adsorbed Ge probe atom within the symmetry-irreducible wedge of each
cell \citep{cha03}. PES data for the whole cells were unfolded by
applying all $C_{3V}$ symmetry operations to the irreducible PES
wedge data. For each \emph{xy} coordinate, the adsorbed atom was initially
placed approximately $3$~Å above the surface, and its \emph{z}-coordinate
was allowed to relax while the coordinates in the \emph{xy}-plane
were kept fixed. The positions of all remaining atoms, but the bottom
Ge-H units, were fully optimized until atomic forces became less than
$0.01$~eV/Å. For $5\times5$ ($7\times7$) supercell, a total of
66 (120) points, forming the irreducible hexagonal grid with $1.2$~Å
spacing, were initially calculated. This method has been successfully
used by several authors (see for instance Refs.~\citep{kaw96,cha03,wan08b,kra13}
for a few works in silicon).

After identification of the main energy minima, additional points
were calculated along the energy paths connecting them, resulting
in a spacing of $0.6$~Å. The exact energies of local energy minima
on PES were calculated with a free-moving probe atom placed near the
local energy minima. Judging from the calculated data, the numerical
uncertainty of calculated energy barriers is estimated at about $0.1$~eV.

\section{RESULTS AND DISCUSSION}

\begin{figure}
\includegraphics[clip,width=8cm]{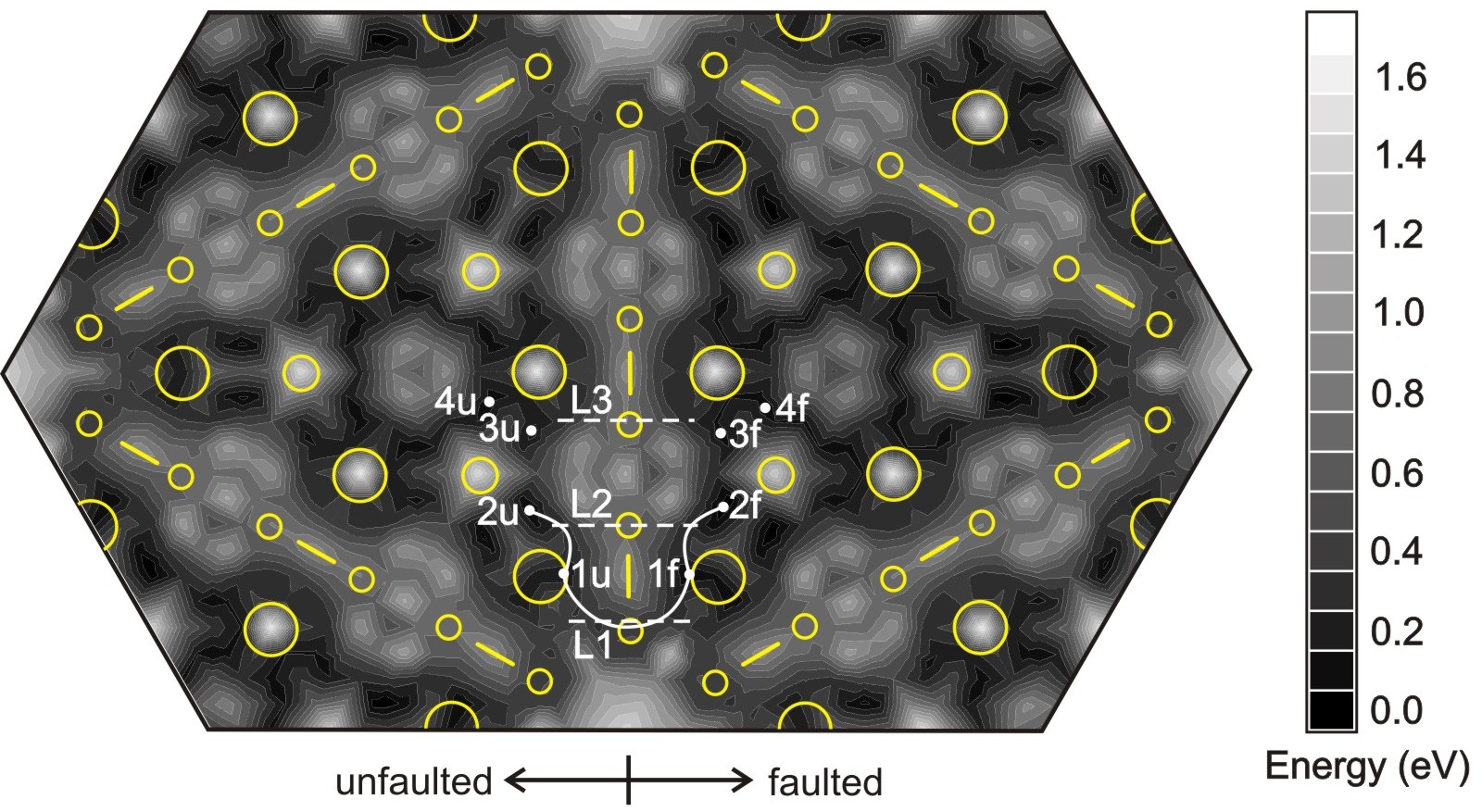}

\caption{\label{fig2} PES for an adsorbed Ge atom on unstrained Ge$(111)\textrm{-}7\times7$
surface. Dark (bright) regions indicate energy minima (maxima) and,
hence, favorable adsorption positions (unstable/transition-state regions).
The contour spacing is $0.1\,\mathrm{eV}$. Large circles point to
adatom locations, medium circles to rest atoms, small circles connected
by rods indicate dimers along the HUC perimeter. White dots with labels
1u-4u and 1f-4f mark the deepest PES minima inside of unfaulted (u)
and faulted (f) HUCs, respectively. Dashed lines across the HUC border
(with labels L1-L3) mark the positions of calculated migration paths.
The solid line connecting the deepest energy minima in faulted and
unfaulted HUCs (2u and 2f) is the minimum energy path for border crossing.}
\end{figure}
\begin{figure*}
\includegraphics[clip,width=12cm]{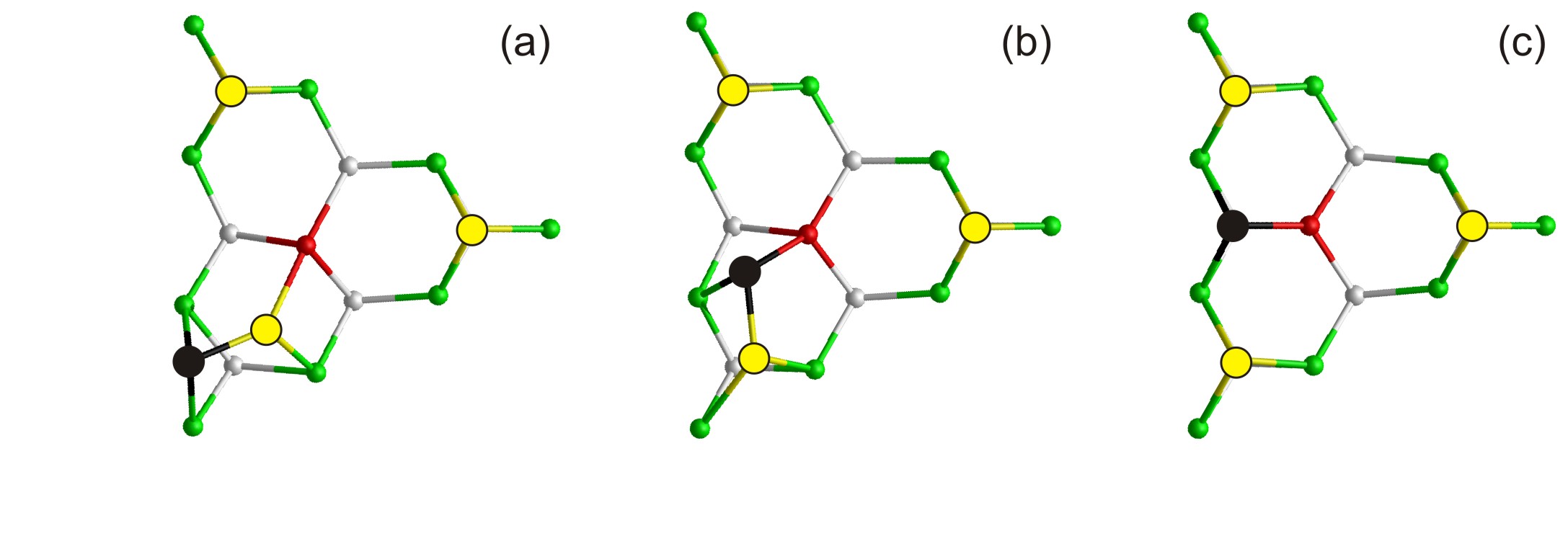}

\caption{\label{fig3} Atomic structures pertaining to the PES minima of Ge
atom on the Ge$(111)\textrm{-}5\times5$ HUC (top view). The coloring
scheme is identical to that of Fig.~\ref{fig1}, with the black ball
representing the adsorbed Ge atom. (a) Unstrained $5\times5$, PES
minima 1u (1f). Same local atomic arrangement is observed for unstrained
$7\times7$, 1u (1f) and strained $5\times5$, 1f. (b) Unstrained
$5\times5$, PES minima 2u (2f). Same local atomic arrangement is
observed for unstrained $7\times7$, 2u (2f), 3u (3f), 4u (4f) and
strained $5\times5$, 2u. (c) Strained $5\times5$, PES minima 5u
(5f) represented in Fig.~\ref{fig5}.}
\end{figure*}

Let us first discuss the Ge/Ge$(111)\textrm{-}7\times7$ system. Figure~\ref{fig2}
shows the calculated PES for the adsorption of a Ge atom on the unstrained
Ge$(111)\textrm{-}7\times7$ surface. Dark (bright) regions indicate
energy minima (maxima). Relative energies of PES minima on Ge$(111)\textrm{-}5\times5$
and Ge$(111)\textrm{-}7\times7$ surfaces are given in Tab.~\ref{tab1}.
From the PES, one can see that the Ge$(111)\textrm{-}7\times7$ PES
is very corrugated. The deepest PES minima are located inside of the
unfaulted (1u-4u) and faulted (1f-4f) HUCs next to adatoms and rest
atoms, where the adsorbed atom can saturate several surface dangling
bonds. From visual inspection of the relaxed coordinates we conclude
that all these PES minima are associated with the formation of dimers
consisting of adsorbed (diffusing) Ge atom and $7\times7$ adatom.
Atomic structures relevant to the PES minima on Ge$(111)\textrm{-}5\times5$
and Ge$(111)\textrm{-}7\times7$ surfaces are shown in Fig.~\ref{fig3}.

Essentially, Figs.~\ref{fig3}(a) and \ref{fig3}(b) show that an
adsorbed Ge atom (black) in strain-free Ge(111), pairs with a surface
adatom (yellow) in the vicinity of a rest atom (red). The resulting
\emph{ad-dimer} sits on top of four surface dangling bonds (one from
the rest atom plus three from backbond atoms) located at the vertices
of a $(111)\textrm{-}1\times1$ primitive cell. Two conspicuously
stable geometries were found for the ad-dimer: (i) structures related
to PES minima 1u (1f), where both atoms of the ad-dimer share a T$_{4}$
site (see Fig.~\ref{fig3}(a)), and (ii) structures related to 2u
(2f) minima, where the ad-dimer is displaced towards the rest atom,
leaving the T$_{4}$ site occupied by a single atom (see Fig.~\ref{fig3}(b)).

Regardless of the configuration adopted, each atom in the ad-dimer
establishes three covalent bonds — one with its dimer partner, plus
two with rest/backbond surface atoms — leaving two dangling bonds
(one on each dimer atom). However, because of the different potential
experienced by each atom, the resulting geometry is asymmetric.

The structure depicted in Fig.~\ref{fig3}(b) is also representative
of PES minima 3u (3f) and 4u (4f), related to the center adatoms (Fig.~\ref{fig2}).
However, according to Tab.~\ref{tab1}, the minima 3u (3f) and 4u
(4f) are less stable than 2u (2f), probably because center adatoms
are more strongly bonded to the substrate compared to corner adatoms,
making them slightly more difficult to shift from their native $\mathrm{T_{4}}$
site. This is consistent with the predicted exclusive formation of
1u (1f) ad-dimers involving corner adatoms, and it is in agreement
with experimental STM results showing that Ge atoms on Si$(111)\textrm{-}7\times7$
surface preferentially substitute Si corner adatoms in faulted HUC
\citep{wan05}.

According to Tab.~\ref{tab1}, sites 2u and 2f are the lowest energy
sites in respectively unfaulted and faulted unstrained HUCs (for both
$7\times7$ and $5\times5$ structures). We note that stable sites
in the faulted HUC have slightly lower energy than the analogous sites
in the unfaulted HUC. The preferential adsorption of adatoms on faulted
Si$(111)\textrm{-}7\times7$ HUCs at low coverage levels has been
previously reported for a variety of atomic species \citep{zha10,tos88,gan91,tan95,lin96,has91,wat98}.
Our results suggest that the same effect should be observed for Ge/Ge$(111)\textrm{-}7\times7$
and Ge/Ge$(111)\textrm{-}5\times5$ at low temperature.

\begin{table}
\begin{ruledtabular}
\caption{\label{tab1}Relative energies (eV) of local stable adsorption sites
for a Ge atom on unstrained Ge$(111)\textrm{-}7\times7$, Ge$(111)\textrm{-}5\times5$
surfaces and Ge$(111)\textrm{-}5\times5$ surface compressively strained
by 4\%. $N$ is the site number according to Figs.~\ref{fig2} and
\ref{fig5}. For each studied surface, the energy of the most stable
site is considered the origin of the energy scale.}
\begin{tabular}{c|cccccc}
\multirow{3}{*}{$N$} & \multicolumn{6}{c}{Surface structure}\tabularnewline
\cline{2-7} \cline{3-7} \cline{4-7} \cline{5-7} \cline{6-7} \cline{7-7} 
 & \multicolumn{2}{c|}{$7\times7$ unstrained} & \multicolumn{2}{c|}{$5\times5$ unstrained} & \multicolumn{2}{c}{$5\times5$ strained}\tabularnewline
\cline{2-7} \cline{3-7} \cline{4-7} \cline{5-7} \cline{6-7} \cline{7-7} 
 & u & \multicolumn{1}{c|}{f} & u & \multicolumn{1}{c|}{f} & u & f\tabularnewline
\hline 
1 & 0.22 & 0.11 & 0.20 & 0.09 & - & 0.19\tabularnewline
2 & 0.07 & 0.00 & 0.07 & 0.00 & 0.18 & -\tabularnewline
3 & 0.15 & 0.12 & - & - & - & -\tabularnewline
4 & 0.09 & 0.11 & - & - & - & -\tabularnewline
5 & - & - & - & - & 0.28 & 0.00\tabularnewline
\end{tabular}
\end{ruledtabular}

\end{table}

Figure~\ref{fig2} clearly shows that PES minima are located around
adatoms and rest atoms, forming a diffusion network within the HUCs
with low energy barriers (about $0.2\textrm{-}0.3$~eV). The reason
for such low energy barriers rests on the distribution of dangling
bonds within the HUCs, favoring a concurrent bond breaking/formation
between the adsorbed atom and substrate atoms along the migration
path.

\begin{figure}
\includegraphics[width=8.5cm]{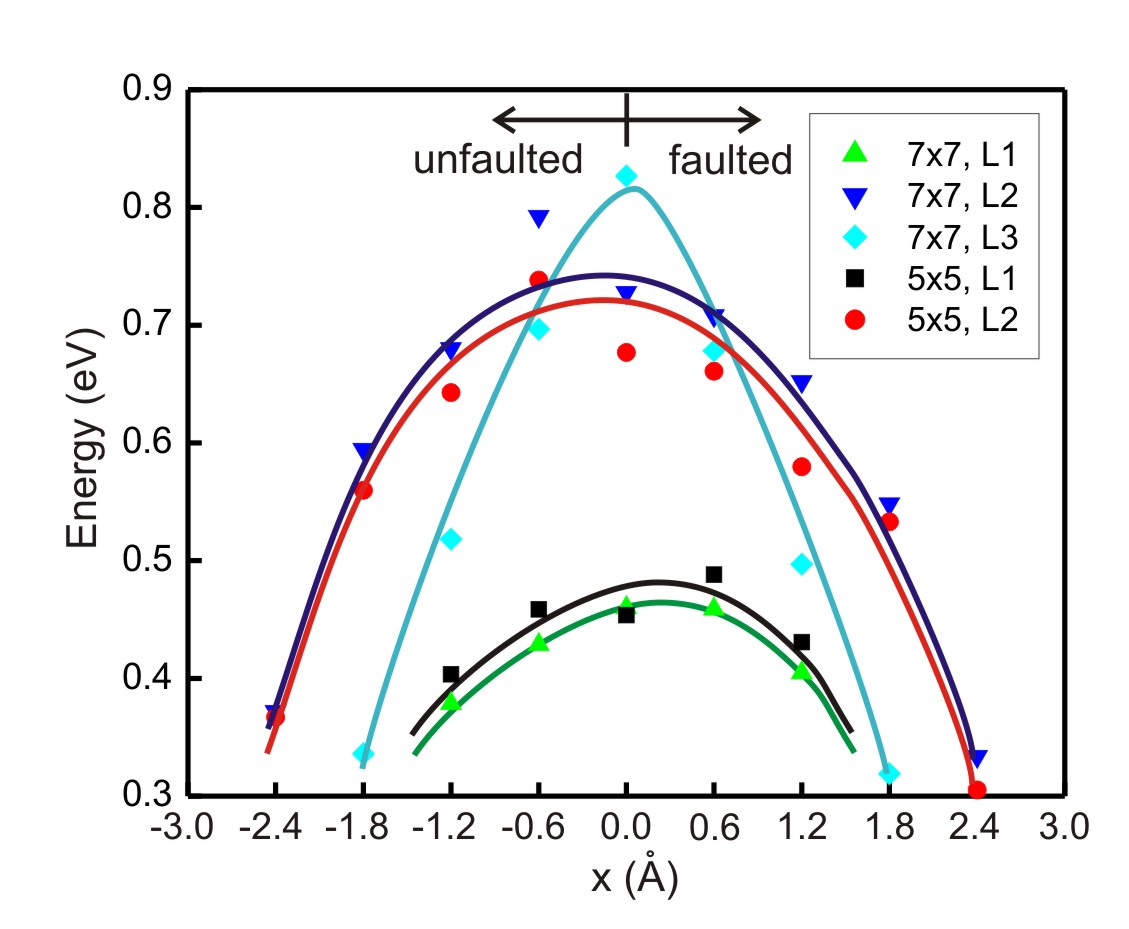}\caption{\label{fig4} Higher-energy section of PES profiles on unstrained
Ge$(111)\textrm{-}7\times7$ (L1-L3 paths) and Ge$(111)\textrm{-}5\times5$
(L1, L2 paths) surfaces as indicated by dashed lines in Figs.~\ref{fig2}
and \ref{fig5}(a) respectively. The energy of the most stable site
(2f) is the origin of the energy scale for all profiles. Data points
are from DFT calculations, solid lines are just guides to the eye.}
\end{figure}

Adjacent HUCs are separated by relatively high energy barriers (bright
regions in Fig.~\ref{fig2}). The reason for the high barriers is
the absence of dangling bonds in regions along the dimer rows. Any
border crossing performed by an adsorbed atom, must break or substantially
weaken its bonds to the substrate. According to the PES, the lowest
energy barriers across HUC border involve a hop of the adsorbed Ge
atom over dimer atoms, either near the CH as indicated by the dashed
line L1, or between dimers (as indicated by dashed lines L2, L3).
Fig.~\ref{fig4} shows the relevant section of three PES profiles
across L1-L3 border crossings depicted in Fig.~\ref{fig2}. For each
profile in Fig.~\ref{fig4} the origin of the energy scale is the
most stable site (2f) of the corresponding surface (either $7\times7$
or $5\times5$).

Clearly, the connection of sites 2u and 2f through L1 is the minimum
energy path (MEP), with the saddle point located near the CH. The
overall energy barrier for hopping from faulted to unfaulted HUCs
along this path is about $0.4\textrm{-}0.5$~eV (Fig.~\ref{fig4}),
and approximately $0.1$~eV lower in the opposite direction (Tab.~\ref{tab1}).
These barriers are distinctively lower than any other barrier for
crossing the HUC border, and as explained below, they determine the
long-range diffusivity of adsorbed Ge atoms across Ge(111)-$7\times7$.
The paths represented by lines L2 and L3 have considerably higher
energy barrier ($0.7\textrm{-}0.8$~eV) and should have limited contribution
to long-range surface diffusion. We note that the MEP on the $7\times7$
reconstructed surface depends on the chemical nature of the adsorbed
atom, most notably on its valence and size. For example, the MEP for
a Sr atom on Si$(111)\textrm{-}7\times7$ crosses the HUC border right
between surface dimers \citep{zha10}.

We suggest that the reason for the relatively low energy barriers,
is formation of a weak bond between the diffusing Ge adatom and dimer
atom. Hybrid $sp^{3}$-states have several lobes responsible for the
formation of strong $\sigma$-bonds. These bonds determine the dimer
structure and are all saturated within $7\times7$ and $5\times5$
structures. However, $sp^{3}$-hybrid states on atoms with bond angles
departing from the perfect tetrahedral geometry, develop wavefunction
lobes with polarization opposite to the connected atoms \citep{bech03}.
Those lobes are prone to form weak covalent bonds, in the present
case with the adsorbed Ge atom. We suggest that such bonds are crucial
in the formation of the energy barriers, pretty much in the same way
as they determine the small energy preference of $\mathrm{T_{4}}$
over $\mathrm{H_{3}}$ adsorption sites for many adsorbate species
on Si$(111)$ \citep{nor84,mea89}. The higher barriers along the
L2 and L3 paths as compared to L1 might be caused by a tight space
(and stronger bonds) between dimers as compared to the space near
the CH, so that formation of bonds with the adsorbed Ge atom are not
so favorable. We also note that the adsorbed Ge atom does not form
a bond with the atom in the middle of the CH — this atom is fully
saturated \citep{ste02}.

The results presented above are in qualitative agreement with DFT
data obtained for the Si/Si$(111)\textrm{-}7\times7$ system \citep{cha03}.
The main difference is that the potential energy landscape for case
of Ge/Ge$(111)\textrm{-}7\times7$ is flatter. For example, the MEP
between HUCs in Si/Si$(111)\textrm{-}7\times7$ also passes near the
CHs and its energy barrier is approximately $1$~eV. Additionally,
energy barriers in the range $0.3\textrm{-}0.7$~eV were found for
motion of adsorbed Si within the faulted and unfaulted HUC areas.
Thus, while at $T=300$~K a Si atom essentially roams within HUC
areas, the crossing between them being hampered by a relatively high
barrier, at $T=400$~K hops between adjacent HUCs were experimentally
observed \citep{sat00}. However, according to our results, the effective
hopping frequency ($\nu$) between HUCs in Ge/Ge$(111)\textrm{-}7\times7$
should be much higher than that in Si/Si$(111)\textrm{-}7\times7$.
Assuming that $\nu$ follows an Arrhenius dependence with $T$, we
readily find that $\nu_{\textrm{Ge}}/\nu_{\textrm{Si}}\approx\exp[(E_{\textrm{Si}}-E_{\textrm{Ge}})/k_{\textrm{B}}T]\approx10^{8}$
at $T=300$~K, considering $E_{\textrm{Si}}=1.0$~eV and $E_{\textrm{Ge}}=0.5$~eV.

The authors of Ref.~\onlinecite{cha03} have shown that migration
via exchange with adatoms within the HUCs of Si/Si$(111)\textrm{-}7\times7$,
also involves a low energy barrier, and therefore should compete with
atomic hopping inside HUC regions. An analogous mechanism may happen
in Ge/Ge$(111)\textrm{-}7\times7$ as well. However, the limiting
step for long-range surface diffusion is HUC border crossing, and
for that reason the exchange with adatoms was not examined. We also
assume that HUC border crossings via exchange with dimer atoms is
highly unlikely. This assumption is based on the fact that all dimer
atoms within the $7\times7$ structure are fully saturated, and any
atomic exchange must involve a costly bond breaking step.

\begin{figure*}
\includegraphics[width=14cm]{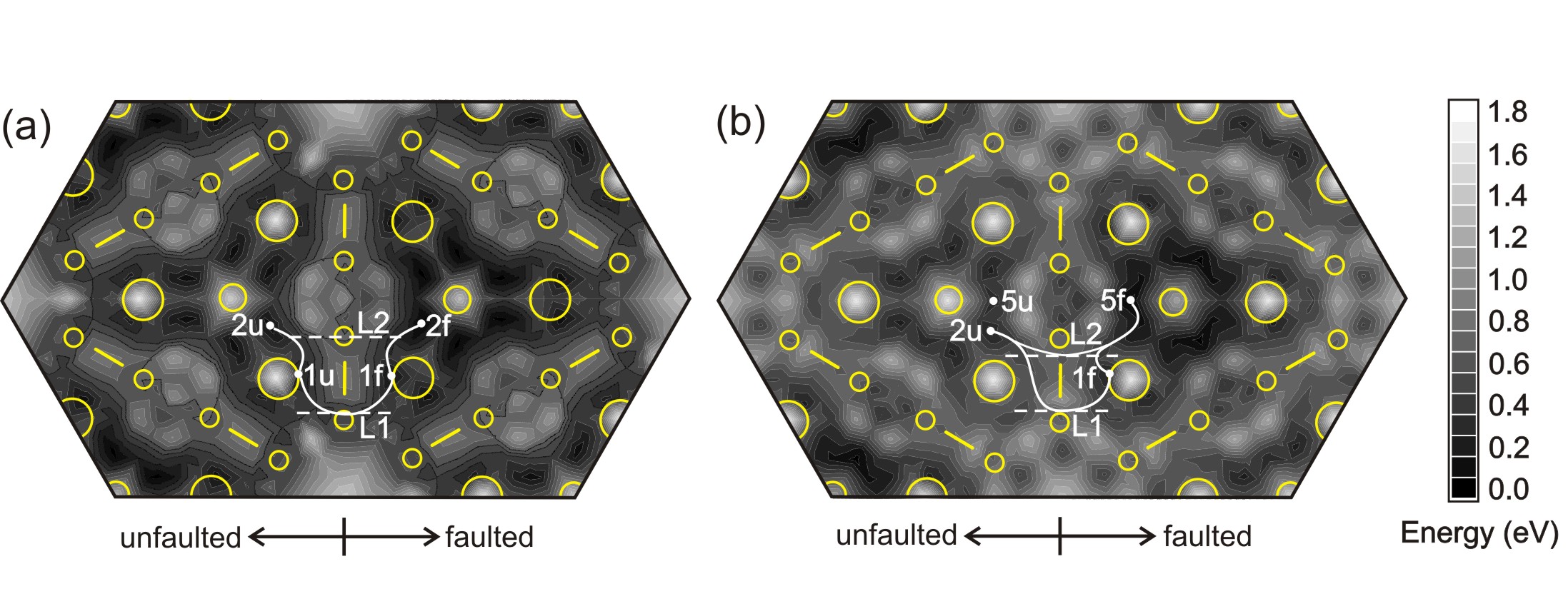}

\caption{\label{fig5}PES for an adsorbed Ge atom on (a) unstrained and (b)
4\% strain-compressed Ge$(111)\textrm{-}5\times5$ surfaces. Dark
(bright) regions indicate energy minima (maxima), and hence, favorable
adsorption positions (unstable/transition-state regions). The contour
spacing is $0.1\,\mathrm{eV}$. Large circles are adatoms, medium
circles are rest atoms, small circles with rods between them are dimers
along HUC perimeter. White dots mark the deepest PES minima inside
of unfaulted (u) and faulted (f) HUCs. Dashed lines across the HUC
perimeter (L1, L2) mark the positions of calculated migration paths.
Solid lines connecting the deepest energy minima (2u and 2f in (a)
and 2u and 5f in (b)) in faulted and unfaulted HUCs are MEPs.}
\end{figure*}

Now we turn to the Ge/Ge$(111)\textrm{-}5\times5$ system. Figure~\ref{fig5}(a)
shows the calculated PES for the adsorbed Ge atom on strain-free Ge$(111)\textrm{-}5\times5$
surface. The image has clear similarities with the PES for Ge/Ge$(111)\textrm{-}7\times7$
presented in Fig.~\ref{fig2}. This is not surprising since both
$5\times5$ and $7\times7$ consist of the same structural elements
— dimers, adatoms and stacking faults — arranged in the same fashion.
The positions that correspond to adatoms at the faulted (unfaulted)
HUCs are dark (bright) indicating that adatoms in the faulted part
are easily shifted from their $\mathrm{T_{4}}$ sites, thus forming
ad-dimer configurations like the corner adatom sites of Ge(111)-$7\times7$.
Again, this is in agreement with experimental STM results \citep{wan05}.

Stable adsorption sites on $5\times5$ are located at 1u (1f) and
2u (2f) minima, with 2u (2f) having the lowest energy (as in case
of $7\times7$). The stable sites on $5\times5$ and $7\times7$ have
identical local atomic arrangement (Figs. 3(a) and (b)). The MEP between
adjacent $5\times5$ HUCs (solid line in Fig.~\ref{fig5}(a)) is
also analogous to that in $7\times7$ HUCs: the crossing point is
close to the CH (Fig.~\ref{fig2}) along the L1 line. The resulting
energy barrier along this MEP, represented by black squares, is very
close to the one obtained for the Ge$(111)\textrm{-}7\times7$ surface
(Fig.~\ref{fig4}). Clearly, the close structural similarity of $5\times5$
and $7\times7$ reconstructions leads to virtually the same energy
barrier that limits long-range surface self-diffusion.

The diffusivity ($D$) describing the macroscopic surface diffusion
of Ge atoms across many $5\times5$ or $7\times7$ cells is given
by $D=D_{0}\exp\left(-E_{\textrm{b}}/k_{\textrm{B}}T\right)$, where
$D_{0}$ is a pre-exponential factor and $E_{\textrm{b}}$ is an effective
energy barrier. Elaborate kinetic Monte Carlo simulations could provide
us with some figures for these quantities. Instead, we consider a
simple surface diffusion model, which has the advantage of directly
addressing the effects presented and questions raised in the Introduction.

The following analysis is based on two assumptions. The first is that
(i) long-range diffusivity is mostly limited by the lowest HUC crossing
barrier. Secondly, since the barriers for hoping within HUC areas
are considerably lower than for HUC border crossing, we also assume
that (ii) adsorbed atoms inside HUCs perform a fast random-walk, which
in the time scale between two consecutive border crossings, allows
them to span the whole HUC area and visit all stable sites with equal
probability.

We start by comparing the surface diffusivity in $7\times7$ and $5\times5$
reconstructed surfaces. According to the above, hops between adjacent
HUCs take place from sites near the CHs along L1 paths only (where
the lowest energy barrier is located). The migrating atom can be described
as effectively occupying the whole HUC area ($S_{\textrm{HUC}}$),
and performing jumps between neighboring HUCs with length $l$ equivalent
to the HUC edge distance. The probability to find the adsorbed atom
at corner sites ready for border crossing, must be $S_{\textrm{cs}}/S_{\textrm{HUC}}$,
where $S_{\textrm{cs}}$ represents an effective surface area of all
corner sites per HUC. Clearly, $S_{\textrm{cs}}$ is the same for
$7\times7$ and $5\times5$ structures, while their jump distance
ratio is $l_{7\times7}/l_{5\times5}=7/5$.

The surface diffusivity of an atom performing independent, randomly
oriented jumps, across a two-dimensional lattice of stable sites,
is generally given by \citep{kon06},

\begin{equation}
D=gl^{2}\nu/4,\label{eq:rate_huc}
\end{equation}
where in the present case $\nu=\nu_{0}\exp(-E_{\textrm{b}}/k_{\textrm{B}}T)$
is the thermal frequency of jumps performed by the adsorbed atom between
adjacent HUCs, $g$ is the number of equivalent hoping routes available
($g=3$ for hexagonal lattice), and $\nu_{0}$ is the frequency of
attempt for the jumps (across the HUC borders). Note that $\nu_{0}$
is different than the attempt frequency for hops within the HUC. The
latter quantity is often approximated to the Debye frequency of the
material, $\nu_{\textrm{D}}$, which to first approximation $\nu_{0}\approx\nu_{\textrm{D}}(S_{\textrm{cs}}/S_{\textrm{HUC}})$
scales with $l^{-2}$.

Considering the above, and taking into account that $E_{\textrm{b}}$
barriers for HUC border crossing are virtually the same in $7\times7$
and $5\times5$ reconstructed Ge$(111)$ surfaces, it follows that
the diffusivity is independent of $l$ and therefore $D_{7\times7}=D_{5\times5}$.
This explains why the effect of surface reconstruction was found negligible
during diffusivity measurements \citep{che04}.

Figure~\ref{fig5}(b) shows the calculated PES for the adsorption
of a Ge atom on a 4\% strain-compressed Ge$(111)\textrm{-}5\times5$
surface along lateral directions (corresponding to the lattice mismatch
between Ge and Si lattice constants). From the above, we trust that
the PES of strained Ge$(111)\textrm{-}7\times7$ surface should lead
to similar results. One can see that the minima 1u and 2f are eliminated,
while new minima – 5u and 5f – emerge. The atomic structure relevant
to 5u (5f) minima is shown in Fig.~\ref{fig3}(c). According to this
figure, the adsorbed Ge atom occupies the $\mathrm{T_{4}}$ site between
two neighboring adatoms. It connects to a rest atom (red) and two
backbond atoms (green). The latter also connect to adatoms (yellow),
and therefore are over-coordinated (have 5 neighbors).

It is interesting to note that while CHs on unstrained surfaces are
clear energy maxima (Figs.~\ref{fig2} and \ref{fig5}(a)), several
metastable sites are formed within these regions upon compression
(Fig.~\ref{fig5}(b)). These sites are separated by large energy
barriers, and they may trap adsorbed Ge atoms at low temperatures.
Considering their locations, we suggest that they are caused by formation
of back-bonds with dimer atoms, which become closer under strain.

\begin{figure}
\includegraphics[width=8.5cm]{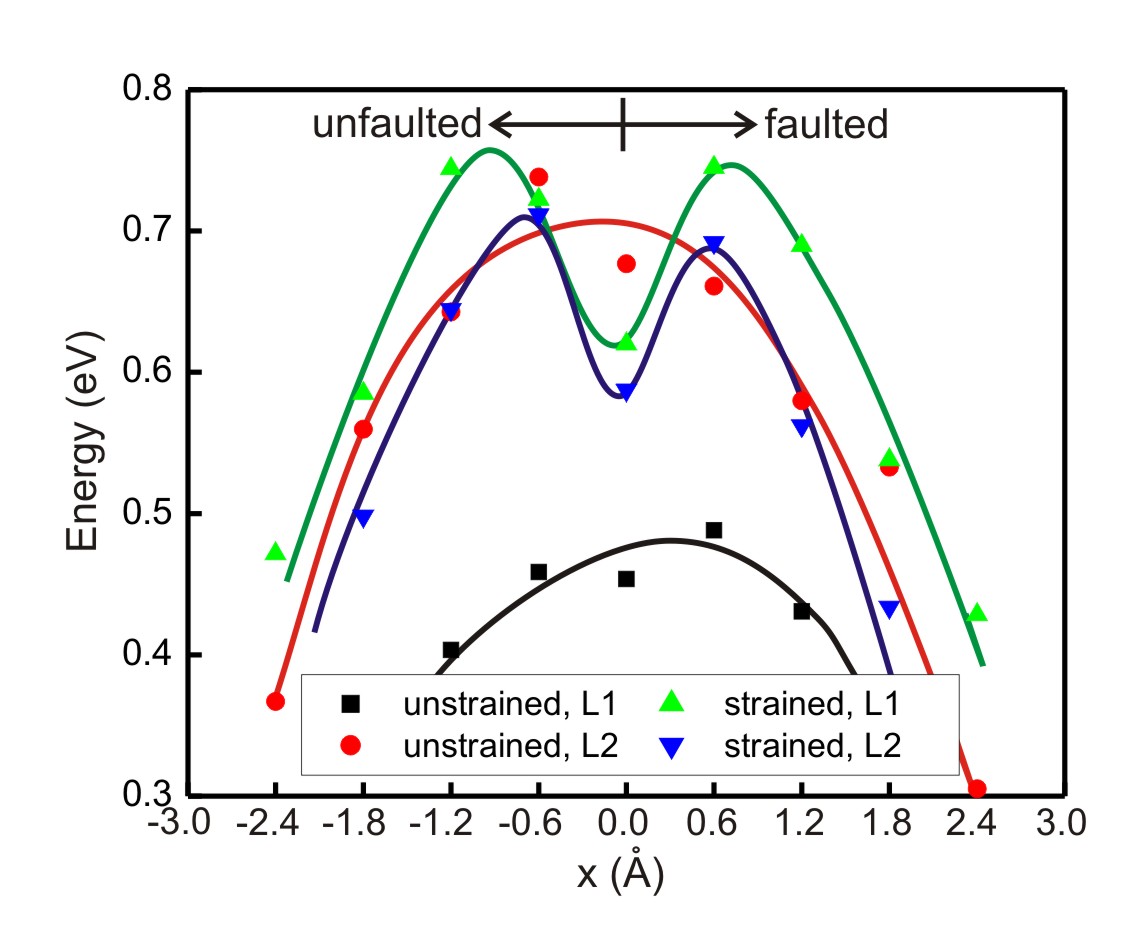}

\caption{\label{fig6}Higher-energy section of PES profiles along L1 and L2
paths on (a) unstrained and (b) 4\% strain-compressed Ge$(111)\textrm{-}5\times5$
surfaces as indicated by dashed lines in Figs.~\ref{fig5}(a) and
(b) respectively. For each studied surface, the energy of the most
stable site is considered the origin of the energy scale (2f for unstrained
and 5f for compressively strained $5\times5$). Data points are from
DFT calculations, solid lines are just guides to the eye.}
\end{figure}

Fig.~\ref{fig6} shows PES profiles along L1 and L2 lines crossing
the HUC border on strain-compressed Ge$(111)\textrm{-}5\times5$ surface.
For each profile in Fig.~\ref{fig6} the origin of the energy scale
is the most stable site of the corresponding surface (2f and 5f for
unstrained and strained $5\times5$, respectively). The profiles have
the shape of a “camel back”, with a metastable midpoint located at
the HUC edge, corresponding to the shallow minima observed near the
CH in Fig.~\ref{fig5}(b). Importantly, the energy barrier along
the L1 line is significantly increased from $0.4\textrm{-}0.5$~eV
on unstrained $5\times5$ to $0.7\textrm{-}0.8$~eV on strained $5\times5$.
On the other hand, the change is not so conspicuous for the L2 profile.
Both paths along L1 and L2 on strained $5\times5$ have similar energy
barriers and both are likely to contribute to long range surface self-diffusion
(see solid lines in Fig.~\ref{fig5}(b)). The impact of the pre-exponential
change to the diffusivity is expected to be minute ($D$ should be
enhanced by a factor of 2 if we account for twice the number of crossing
points under strain). On the other hand, a change of $E_{\textrm{b}}$
from 0.5~eV to 0.8~eV effectively hinders the adsorbed Ge atom within
the HUC boundaries by decreasing $D$ by a factor of $10^{-5}$ at
$T=300$~K.

Although the experimental data also suggest that the barrier increases
with compression, the estimated change in the energy barrier ($50\textrm{-}70$~meV)
\citep{che02,che04} is less dramatic than the one anticipated theoretically.
Much of the difference could be due to the fact that during the experiments
the strain applied to the Ge$(111)\textrm{-}7\times7$ surfaces was
less than 4\%. Indeed, previous calculations have shown that the $7\times7$
structure of Ge$(111)$ surface is stable in the $\approx0\textrm{-}1\%$
compressive strain range \citep{zha13}. Larger strain fields should
convert the $7\times7$ into the $5\times5$ structure \citep{zha13}.
We suggest that the strain applied to the thin Ge(111)-$7\times7$
layers reported in Refs.~\citep{che02,che04} was below 1\%.

The increase of the energy barrier along L1 line (near the CH) on
compressively strained surface can be caused by formation of stronger
bonds at the 5f global PES minima (Fig.~\ref{fig5}(b)), a destabilization
effect at the saddle point, or both. To clarify this aspect, we compared
the adsorption energy for (i) global PES minima sites and (ii) PES
saddle points along L1 lines on unstrained and strained Ge$(111)$
surfaces. We found that the adsorption energy on PES minima sites
is almost identical for both surfaces. On the contrary, the adsorption
energy on L1 PES saddle point for strained surface, is lower than
for the unstrained one by $\approx0.27$~eV. Therefore, the strain-induced
enhancement of the barrier arises from destabilization of the PES
saddle point.

The assumption on the nature of the weak bond between adsorbed Ge
and dimer atoms made above can also explain the mechanism for bond
weakening at the saddle-point under compression. It is known that
surface dimer bonds are stretched as compared to the bonds in the
crystal bulk (our calculations give 2.617~Å for the length of Ge$(111)\textrm{-}5\times5$
dimer bonds against 2.510~Å for bulk bonds), which is in agreement
with our previous calculations \citep{dol21}. When the surface is
4\% compressed, the dimer bond lengths decrease towards the bulk value
and become more stable (2.576~Å). A stronger $\sigma$-bond means
that more electron density is accumulated between dimer atoms and
less density remains to share with a back-bonded the adsorbed Ge atom.
The results obtained for Ge$(111)\textrm{-}5\times5$ with Ge atom
adsorbed at the saddle point support our arguments: accordingly, we
find 2.516/2.494~Å dimer bond lengths and 2.513/2.606~Å bonds between
adsorbed Ge and dimer atom for relaxed/compressed surface, respectively.

We finally note that the bonds connecting structure adatoms to surface
backbond atoms in Ge(111)-$5\times5$, become stronger under surface
compressive strain. This is evident from the bright spots at the corresponding
sites of faulted HUC in Fig.~\ref{fig5}(b) as compared to dark spots
at these sites in Fig.~\ref{fig5}(a), suggesting that the structure
adatoms become less prone to be displaced from their equilibrium positions.

\section{CONCLUSIONS}

We investigated the effects of surface reconstruction ($5\times5$
or $7\times7$) and strain on diffusion of adsorbed Ge atoms on Ge$(111)$.
The work was accomplished by accurately mapping the potential energy
surface using first principles calculations. According to the proposed
model of surface diffusion, the adsorbed Ge atoms are very mobile
within the half unit cell (HUC) areas, while long-range diffusivity
is limited by a relatively larger barrier for crossing the rows of
dimers (next to corner holes) which separate adjacent HUCs. We found
that the change of surface reconstruction from $7\times7$ to $5\times5$
has negligible effect on long range surface diffusion of Ge adsorbed
atoms. This finding follows from identical energy barriers and atomistic
mechanisms obtained for both $5\times5$ and $7\times7$ structures.
We also found that surface compression strengthens the dimer bonds.
Concurrently, that weakens the bond between dimer and adsorbed Ge
atom at the saddle point, hence increasing the migration barrier of
adsorbed Ge atoms across HUCs.
\begin{acknowledgments}
R. A. Z. would like to thank the Novosibirsk State University for
providing the computational resources. Cluster computations and paper
writing were supported by the Russian Science Foundation (project
no. 19-72-30023). J. C. acknowledges the FCT through projects UIDB/50025/2020-2023
and LA/0037/2020.
\end{acknowledgments}

\bibliographystyle{apsrev4-1}
%\bibliography{refs}


\end{document}